\documentclass[10pt,letterpaper,twocolumn]{article} 

\usepackage{ol}
\usepackage{amsmath}
\usepackage{graphicx}
\begin{document}

\twocolumn[ 

\title{\text Threshold power reduction for forming bright solitons guided in nonlocal nonlinear media}

\author{YuanYao Lin and Ray-Kuang Lee}
\address{Institute of Photonics, National Tsing-Hua University, 101, Section 2, Kuang-Fu Road, Hsinchu City 300, Taiwan}

\begin{abstract}
We study the formation of dark-bright vector soliton pairs in nonlocal Kerr-type nonlinear medium.
We show, by analytical analysis and direct numerical calculation, that in addition to stabilize vector soliton pairs  nonlocal nonlinearity also helps to reduce the threshold power for forming a guided bright soliton.
With help of the nonlocality, it is expected that the observation of dark-bright vector soliton pairs in experiments becomes more workable.
\end{abstract} 

 ] 

Recently the study of nonlocal nonlinearity brings new features in solitons \cite{Snyder97}, such as modification of modulation instability \cite{Krolikowski04} and azimuthal instability \cite{Anton06}, suppression of collapse in multidimensional solitons\cite{Bang02}, change of the soliton interaction \cite{Peccianti02}, and formation of soliton bound states \cite{Torner05}.
Nonlocal effect comes to play an important role as the characteristic response function of the medium is comparable to the transverse content of the wave packet. 
Experimental observations of nonlocal response also have been demonstrated in various systems, such as photorefractive crystals \cite{Duree93}, nematic liquid crystals \cite{Conti03}, thermo-optical materials \cite{Rotschild05}, and $^{52}Cr$ Bose-Einstein condensates with strong dipole-dipole interaction \cite{Pfau05}.

For nonlinear local media, dark-dark, bright-bright, or dark-bright soliton pairs can exist in normal or anomalous dispersion region in vector settings in contrast to the scalar models \cite{Christo88, Kivshar03}.
Especially with the help of a dark soliton, a bright soliton in normal dispersive media is formed through soliton-induced guiding effect \cite{Sheppard97}.
With nonlocal nonlinearity, a large number of multi-hump multi-component vector solitons are found with a remarkable stabilization \cite{Torner06}.
For nonlocal Kerr-type nonlinear medium, the nonlocality is known to improve the stabilization of solitons due to the diffusion mechanism of the nonlinearity.
The price to pay is that nonlocal solitons also need to increase their formation power to compensate the diffusion effect in nonlocal materials.
In this Letter, we study vector solitons with dark-bright pairs in nonlocal nonlinear media.
We reveal that stable bright solitons can be formed in normal dispersive region guided by dark soliton backgrounds.
Moreover we find that the nonlocal nonlinearity also helps to reduce the threshold power for such a guided bright soliton due to the combination of nonlocality and vectorial coupling.
\begin{figure}[hb]
\includegraphics[width=4.15cm]{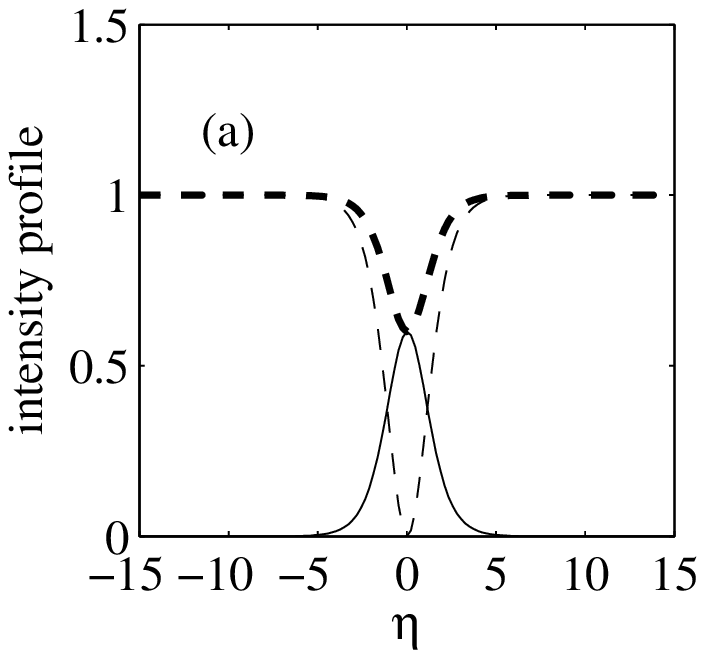}
\includegraphics[width=4.15cm]{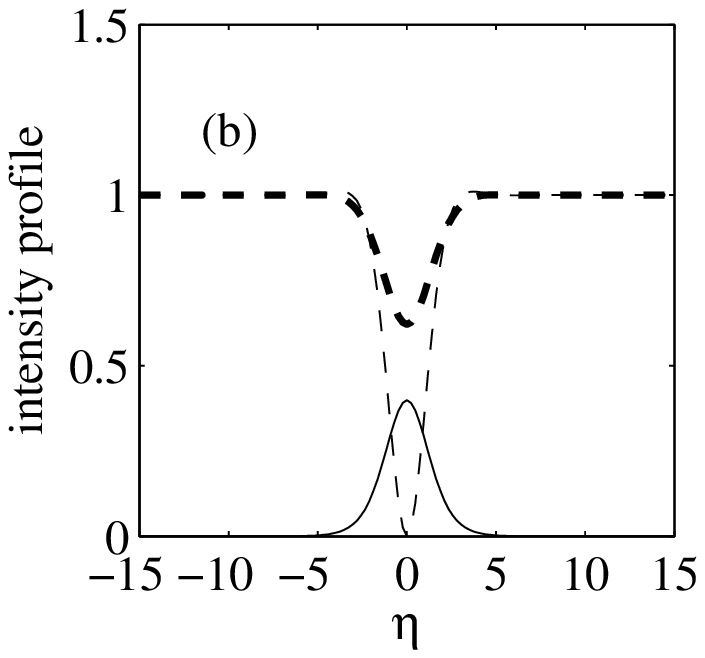}
\includegraphics[width=8.3cm]{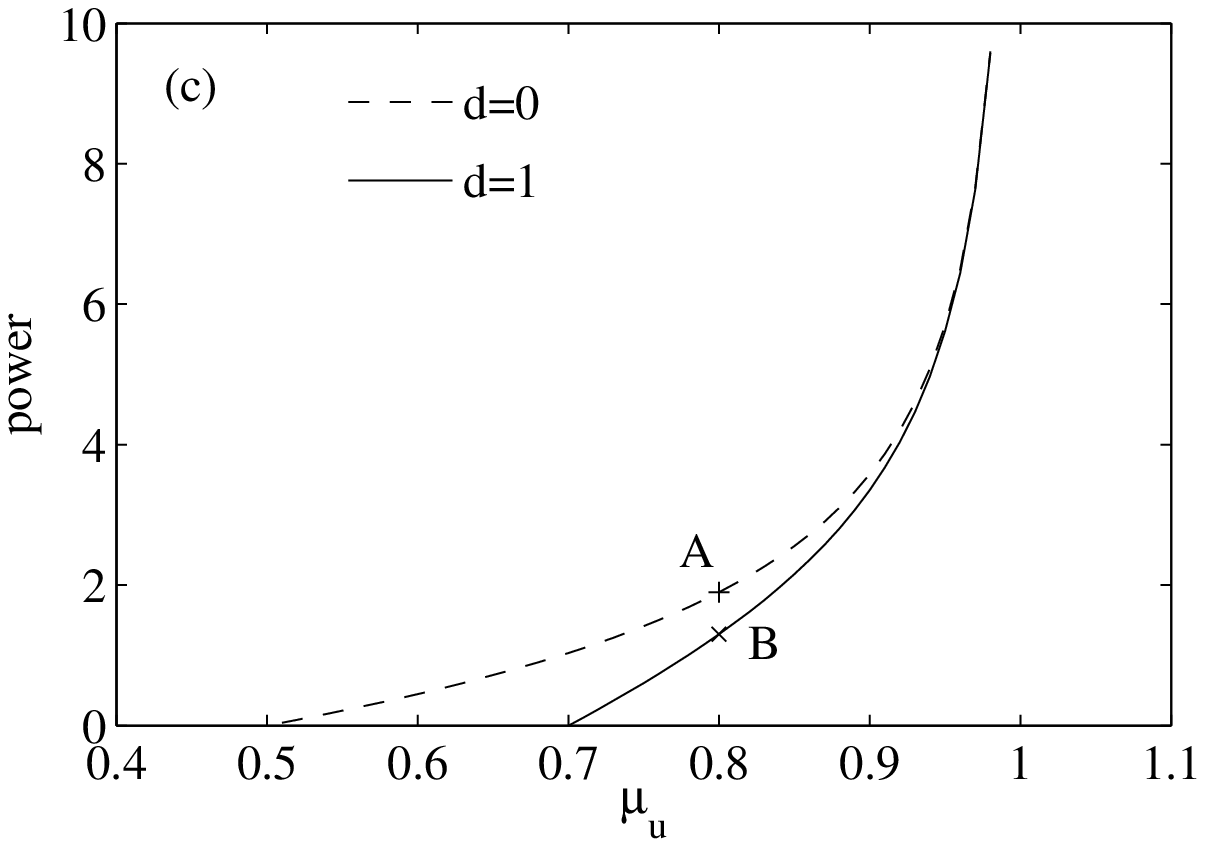}
\caption{Intensity profiles of bright soliton (solid line), dark soliton (dashed line), and the refractive index (bold-dashed line) for the local (a) and nonlocal media (b) at the points $A$ and $B$ in (c), respectively. The bifurcation curves of fundamental dark-bright soliton pairs in local ($d = 0$, dashed line) and nonlocal media ($d = 1$, solid line) are shown in (c), where points $A$ and $B$ are marked with $\mu_u=0.8$ and $\mu_v=1$. }
\label{Fig:F1}
\end{figure}

We consider two mutually incoherent wave packet propagating along the $\xi$ axis within a nonlocal Kerr-type nonlinear medium.
The governing equations of the vectorial Manakov system which consists of two vector components $U$ and $V$ are given by,
\begin{eqnarray}
\label{equ}
&& i \frac{\partial U}{\partial \xi}-\frac{1}{2}\frac{\partial^2 U}{\partial \eta^2}+ n(\xi,\eta) U=0,\\
\label{eqv}
&& i \frac{\partial V}{\partial \xi}-\frac{1}{2}\frac{\partial^2 V}{\partial \eta^2}+ n(\xi,\eta) V=0,\\
\label{eqn}
&& n(\xi,\eta) =\int_{-\infty}^{\infty} R(\eta-\eta')(|U|^2+|V|^2)\text{d} \eta', \\
\label{eqr}
&& R(\eta)=\frac{1}{2 \sqrt{d}}\,e^{-|\eta|/\sqrt{d}},
\end{eqnarray}
where $\eta$ is the transverse coordinates, $n(\xi,\eta)$ is the refractive index profile induced by the exponential-type diffusion kernel function $R(\eta)$ responding to the intensity soliton intensity \cite{Kr2001}.
The coefficient $d$ stands for the degree of nonlocality which governs the diffusion strength of refractive index.
With the nonlocal vector model in Eq. (\ref{equ}-\ref{eqr}), stationary solutions in form of $ V(\xi,\eta)=v(\eta) \text{exp}(i \mu_v \xi)$ and $ U(\xi,\eta)= u(\eta)\text{exp}(i \mu_u \xi)$ are assumed to be the solution of dark-bright vector soliton pairs with real propagation constants, $\mu_v$ and $\mu_u$, respectively.
The two component solutions of dark-bright vector soliton pairs are subject to the boundary condition $u(\pm \infty) = 0$ and $v(\pm \infty) = \pm \sqrt{\mu_v}$.

\begin{figure}
\centerline{\includegraphics[width=8.3cm]{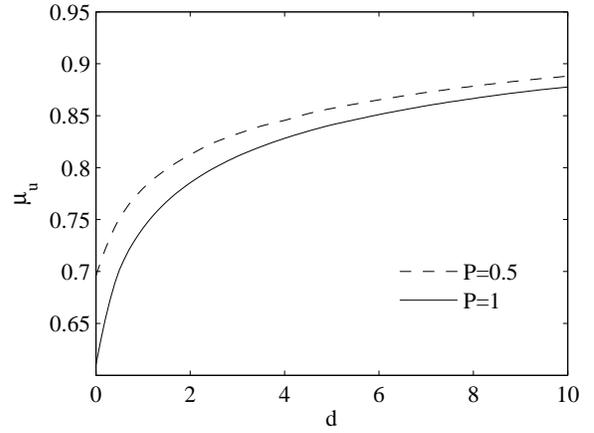}}
\caption{Relations between the propagation constant for bright soliton $\mu_u$ and the degree of nonlocality $d$ at different fixed powers, $P = 0.5$ (solid line) and $P = 1.0$ (dashed line).}
\label{Fig:F2}
\end{figure}

The solutions of dark-bright vector soliton pairs in local and nonlocal media, as well as the refractive index profiles, are shown in Figure \ref{Fig:F1} (a) and (b).
The dependence of the power for bright component, defined in Eq.(\ref{eqpw}), and its propagation constant $\mu_u$ is plotted in Fig. \ref{Fig:F1} (c).
As the case in scalar model, a bright soliton in nonlocal media has higher cutoff potential than the local one due to the diffusion of the nonlinear index.
In Figure \ref{Fig:F2}, the relations between the propagation constant for bright soliton $\mu_u$ and the degree of nonlocality $d$ at different fixed powers are shown.
It can be seen that the propagation constant of the bright one in a dark-bright soliton system is growing as the degree of nonlocality increases, for the bright component sees a potential directly from the nonlinear index that is raised due to the diffusion nonlocality in response to the intensity sum of the dark-bright soliton pair.
But in contrast to the scalar soliton in local media \cite{Kr2000}, the bright soliton guided by a dark soliton in vector model requires lower forming power in nonlocal region as the same propagation constant is concerned, as the marked points $A$ and $B$ shown in Fig. \ref{Fig:F1} (c).

\begin{figure}[t]
\centerline{\includegraphics[width=8.3cm]{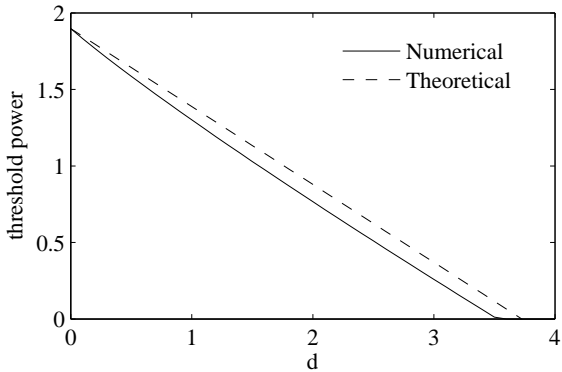}}
\caption{Threshold power of bright solitons versus the degree of nonlocality for $\mu_u = 0.8$ and $\mu_v = 1.0$. Solid and dashed lines are calculated by numerical and variational methods, respectively. }
\label{Fig:F3}
\end{figure}

In addition to direct numerical treatments, we examine the formation power of bright soliton in nonlocal media analytically by variational methods.
To simplify the analysis, a constant dark pulse is assumed as the background and the Lagrangian equation for the bright soliton in a vectorial nonlinear nonlocal system has the form,
\begin{align}
\label{eqLg}
&L = \int\text{d}\eta\{ \frac{i}{2}(U_\xi U^\ast - U_\xi^\ast U ) +  \frac{1}{2}|U_\eta|^2+\frac{1}{2}|U|^4+|U|^2|V|^2 \nonumber \\
&+ d \left[ |U_\eta|^2 +\frac{1}{2}|U|^2(U_{\eta\eta}\,U^\ast + U_{\eta\eta}^\ast\,U)+|U|^2 \frac{\partial^2}{\partial \eta^2}|V|^2  \right]\},
\end{align}
where the subscriptions $\xi$ and $\eta$ stand for derivative with respect to longitudinal and transverse coordinates. 
The terms within the brackets multiplied by $d$ in the second line in Eq. (\ref{eqLg}) represent the nonlocal index response which can be calculated through following expansion,
\begin{eqnarray}
n(\eta) &=& \sum_{m=0}^{\infty} \frac{1}{m!}h_m \frac{\partial^m |U|^2}{\partial \eta^m}\\
&\approx& |U|^2+|V|^2 + d \left[\frac{\partial^2}{\partial \eta^2}(|U|^2+|V|^2) \right],  \nonumber 
\end{eqnarray}
where
\[
h_m = i^m  \frac{\text{d}^m}{\text{d}\, \omega^m} H(\omega)|_{\omega =0},
\]
is the expansion of the Fourier transform, $H(\omega)$, of the kernel function $R(\eta)$ in Eq. (\ref{eqr}).
To solve the Lagrangian equation in Eq. (\ref{eqLg}), we use following solution ansatz for the bright and dark solitons,
\begin{eqnarray}
U(\eta) &=& A_u sech(\eta/a_u)exp(i \phi_u + i c_u \eta^2),\\
V(\eta) &=& A_v tanh(\eta/a_v)exp(i \phi_v + i c_v \eta^2),
\end{eqnarray}
where the parameters $A_j$ , $a_j$, $\phi_j$ and $c_j$, $(j = U, V)$ are amplitude, width, phase, and chirp for bright and dark solitons, separately.
By assuming that dark soliton is invariant to the change of the degree of nonlocality in the low nonlinear limit, a set of Euler-Lagrangian equations for $A_u$, $a_u$,$\phi_u$ and $c_j$ can be obtained.
Furthermore, we assume that in steady state, $\dot{\phi_u}=\mu_u$ is a constant, and $c_j$ is zero for chirpless soliton solutions.
Then for a set of propagation constants of dark-bright soliton pairs, $\mu_v$ and $\mu_u$, an approximate linear dependence of bright soliton power versus nonlocality is derived as,
\begin{align}
\label{eqpw}
P & \equiv \int_{-\infty}^{\infty} |U|^2 \text{d}\eta = 2 A_u^2 a_u, \\ 
& \approx \frac{4 \mu_u -2 \mu_v}{\sqrt{2 \mu_v - 2 \mu_u}} 
+d \left[ -\frac{0.13}{2 \mu_v - 2 \mu_u}+\frac{0.877}{\sqrt{2 \mu_v - 2 \mu_u}}-1.57 \right].\nonumber
\end{align}
In this first-order approximation, Eq. (\ref{eqpw}) indicates that the forming power of bright soliton guided by a dark soliton in vector model decreases as the degree of the nonlocality increases.
In addition, when $d = 0$, Eq. (\ref{eqpw}) reduces to the case of soliton solutions in the local media \cite{Yang01}.
In Fig. \ref{Fig:F3}, we show the dependence of threshold power for forming bright solitons with the degree of nonlocality both by direct numerical simulation of Eq. (\ref{equ}-\ref{eqr}) and the Lagrangian equation in Eq. (\ref{eqLg}), which is consistent with the results in Fig. \ref{Fig:F1}(c).
In this case we fix $\mu_v$ to $1$ since it is associated with the dark component which is given by the boundary conditions and can be scaled out.
Conceptually, as the degree of nonlocality increases, the tendency for refractive index to advance to the region of lower light intensity grows stronger.
Even though the dark pulse almost remains unchanged, the existence of bright pulse drives out index flow.
Consequently the index modulation induced by the soliton pair becomes shallower, and the nonlinearity required to form the bright soliton decreases.
The power of bright soliton decreases in a dynamical balance with the refractive index flow.
This implies that the threshold power to form a bright soliton guided in the dark background can be reduced with nonlocal interaction. 

The stability of dark-bright soliton pairs in nonlocal nonlinear media is analyzed by standard linear stability analysis with introduction of perturbations solutions, i.e.
\begin{eqnarray*}
U &=& e^{i \mu_u \xi}[u_0+(p_u+i q_u)e^{\lambda \xi}+(p_u^\ast+i q_u^\ast)e^{\lambda^\ast \xi}],\\
V &=& e^{i \mu_v \xi}[v_0+(p_v+i q_v)e^{\lambda \xi}+(p_v^\ast+i q_v^\ast)e^{\lambda^\ast \xi}],
\end{eqnarray*}
where the small perturbations grow at the rate of the real part of $\lambda$.
As the case in local media \cite{MLisak1990}, dark-bright soliton pairs in Manakov model are stable in the nonlocal nonlinear media.

In conclusion, we study the formation of dark-bright soliton pairs in vectorial nonlocal nonlinear model analytically and numerically.
We find that in addition to the stabilization of vector soliton pairs, nonlocal nonlinearity also helps to reduce the threshold power for forming a guided bright soliton due to the dynamical balance between the nonlinearity and nonlocal induced refractive index flow.
With a constant background of dark solitons, our analytical model shows a linear dependence of the formation power for bright solitons on the degree of nonlocality and also matches the numerical simulations very well.
With the reduction of forming threshold power, we believe that our results are very useful for the observation of dark-bright vector soliton pairs in nonlocal nonlinear media.

Authors are indebted to Yu. S. Kivshar, W. Kr{\'o}likowski, O. Bang, and A. S. Desyatnikov for useful discussions.
This work is supported by the National Science Council of Taiwan with the contrast number NSC-95-2120-M-001-006.

\end{document}